\documentclass[11pt,twoside]{article}
\usepackage{asp2010}

\resetcounters

\begin{document}

\title{Bonsai: A GPU Tree-Code}
\author{Jeroen B\'{e}dorf$^1$, Evghenii Gaburov$^{1,2}$, and Simon Portegies Zwart$^1$
\affil{$^1$Leiden Observatory, Leiden University, P.O. Box 9513, 2300 RA Leiden, 
The Netherlands}
\affil{$^2$Northwestern University, 2131 Tech Drive, Evanston 60208, IL, USA}
}

\begin{abstract}
  We present a gravitational hierarchical $N$-body code that is designed 
  to run efficiently on Graphics Processing Units (GPUs). All parts of the
  algorithm are exectued on the GPU which eliminates the need for data transfer 
  between the Central Processing Unit (CPU) and the GPU.  
  Our tests indicate that the gravitational 
  tree-code outperforms tuned CPU code for all parts of the algorithm and 
  show an overall performance improvement of more than a factor 20, resulting in 
  a processing rate of more than $2.8$ million particles per second.
\end{abstract}

\section{Introduction}\label{sect:introduction}

A populair method for simulating gravitational $N$-body systems is the 
hierarchical tree-code algorithm orginaly introduced by \cite{1986Natur.324..446B}.
This method reduces the computational complexity of the simulation from ${\cal O}(N^2)$
to ${\cal O}(N\log N)$ per crossing time. The former, though computationally 
expensive, can easily be implemented in parallel for many particles.
The later requires more
communication and book keeping when developing a parallel
method. Still for large number of particles ($N \gtrsim 10^5$)
hierarchical\footnote{ Tree data-structures are commonly referred to
  as hierarchical data-structures. In this work we use an octree data-structure.}
methods are more efficient than
brute force methods. Currently parallel octree implementations are found in a wide
range of problems, such as: self gravitating systems, smoothed
particle hydrodynamics, molecular dynamics, clump finding, ray tracing
and voxel rendering. All of these problems require a high amount 
of computation time. For high
resolution simulations ($N \gtrsim 10^5$) 1 Central Processing Unit 
(CPU) is not sufficient, one has to use computer clusters or
even supercomputers, both of which are expensive and scarce. 
A GPU provides an atractive alternative to such systems.

The GPU is a massively parallel processor which is specialised in
performing independend parallel computations. For an overview
and more details see \cite{2007NewA...12..641P,  2007astro.ph..3100H, 
2008NewA...13..103B, Gaburov2009630}. In this work
we have implemented all the seperate parts of the Barnes-Hut 
tree-code algorithm on the GPU. This includes the tree-construction
and computation of the tree-properties (multipole moments). By doing so we remove the
need to communicate large amounts of data between the CPU and GPU.
Since there is no time lost by CPU-GPU communication we can 
make optimal use of the GPU in a block time-step algorithm.
In previous work \cite{OctGravICCS10,Hamada:2010:TAN:1884643.1884644}
parts of the algorithm were executed on the CPU which only allows 
for efficient (parallel) execution if shared time-steps are used.

Full implementation details, performance and accuracy characteristics 
can be found in \cite{2011arXiv1106.1900B}

\section{Results}\label{Sect:Results}

\begin{figure}
\center \includegraphics[width=0.75\columnwidth]{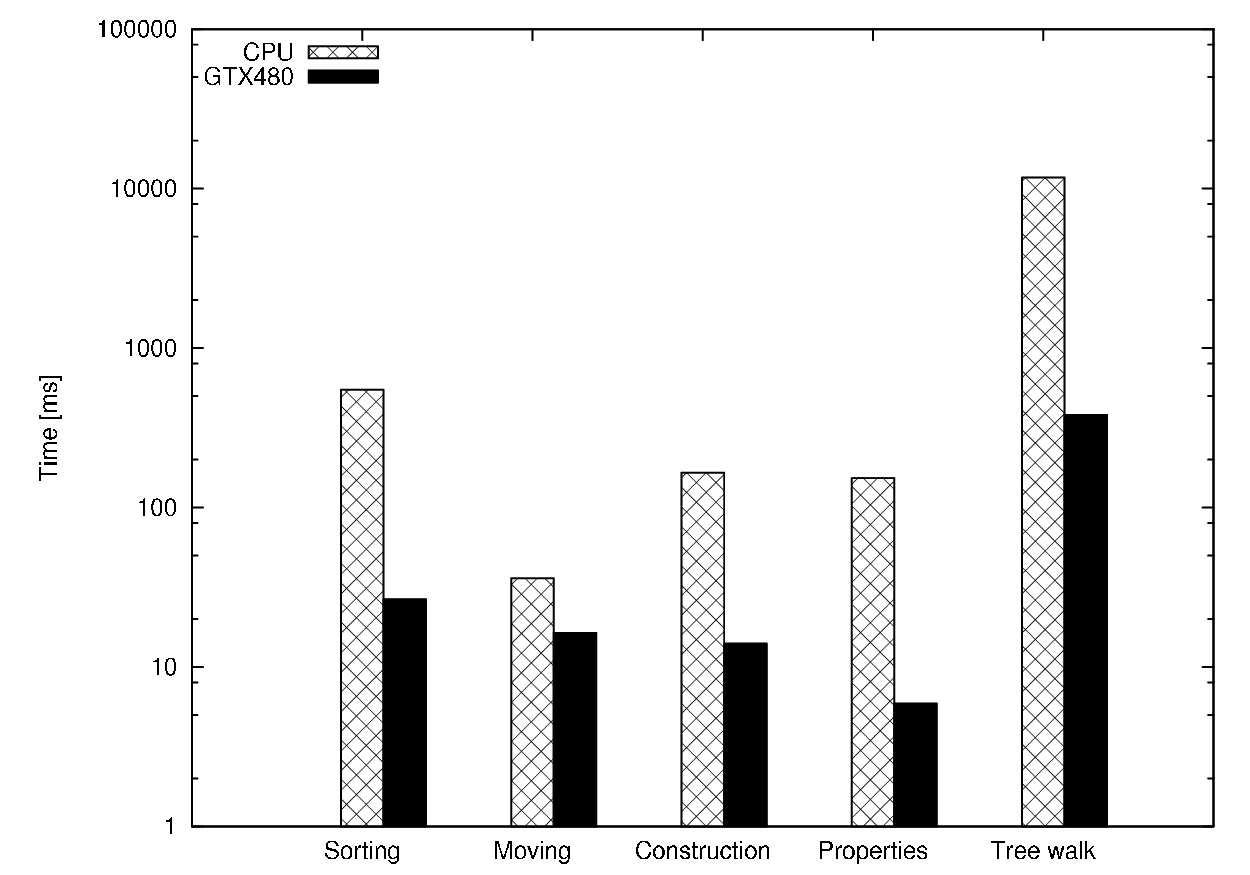}
\caption{Wall-clock time spent by the CPU and the GPU
  on various primitive algorithms. The bars show the time spent
  on the five selected sections of code. The results indicate that our GPU code
  outperforms the CPU code on all fronts and is between 2 and 30 times faster.
  Note that the y-axis is in logscale.
 (Timings using a $2^{20}$ million body Plummer sphere with $\theta=0.75$)}
\label{Image:GenScaling}
\end{figure}

The algorithms are implemented as part of the gravitational $N$-body 
code {\tt Bonsai\footnote{The code is publicly available at: \newline {\tt
      http://castle.strw.leidenuniv.nl/software.html}}}. We compare the 
performance of the GPU
algorithms with (optimized) CPU implementations of comparable algorithms. The 
performance of the tree-traverse depends critically on the multipole acceptance
critera (MAC) ($\theta$) which sets the trade-off between 
speed and accuracy. In our implementation we use  a combination of the
method introduced by Barnes (1994) and the method used for 
tree-traversion on vector machines see \cite{Barnes1994,1990JCoPh..87..161B}.
To compare the CPU and GPU implementations we measure the wall-clock time for the 
most time critical parts of the algorithms.  For the tree-construction we distinguish
three parts; Sorting of the particles along a Space Filling Curve (\cite{Morton}) (sorting in the figure) ,
reordering of particle properties based on the sort (moving) and 
construction of the tree-datastructure (tree-construction). 
Furthermore, timings are presented for the computation of the 
multipole moments and tree-traverse.
As for hardware we used a Xeon E5620 CPU with 4 physical cores and a NVIDIA GTX480 GPU. 
The resuls are presented in Fig.~\ref{Image:GenScaling}.

\begin{figure}
\center
\includegraphics[width=0.75\columnwidth]{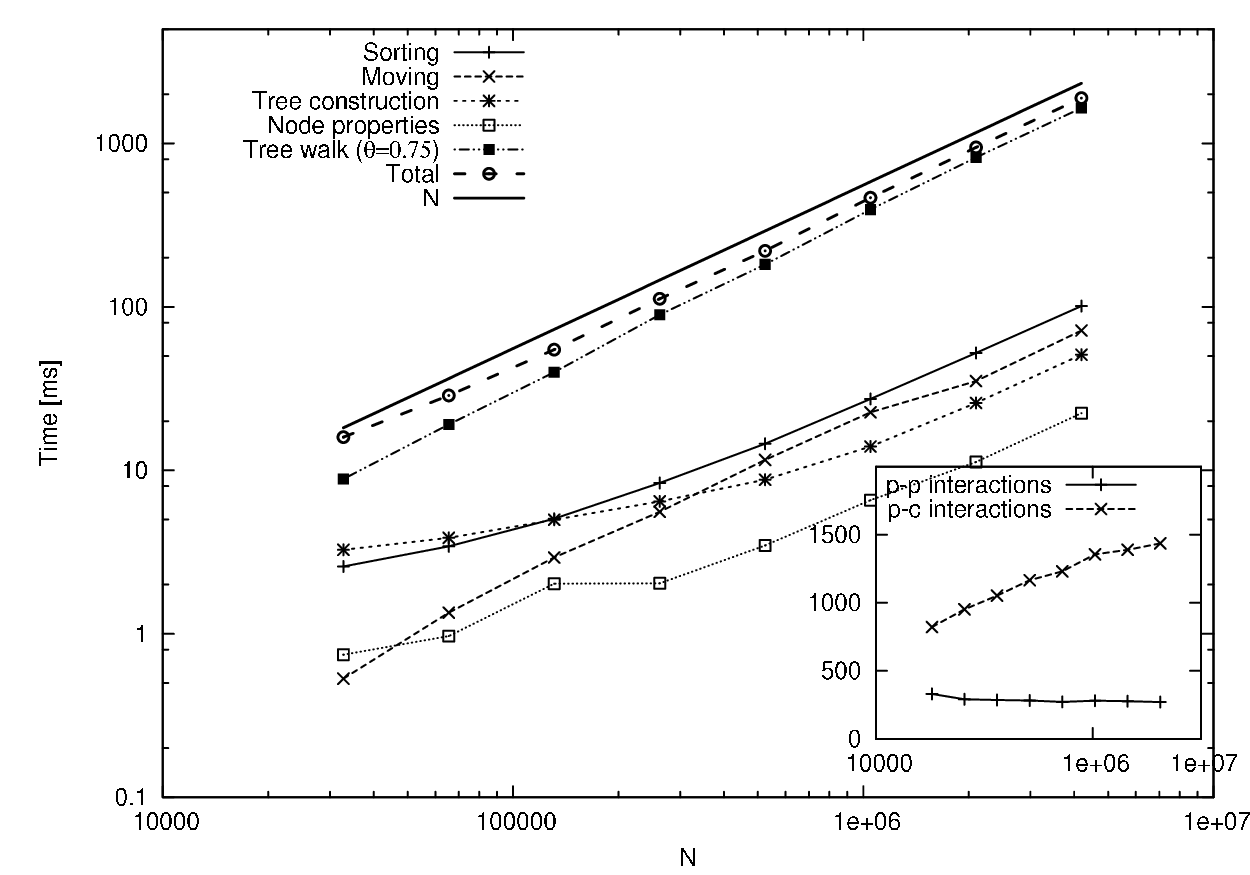}
\caption{The wall-clock time spent by various parts of the
  program versus the number of particles $N$. We used Plummer models as initial conditions
  and varied the number of particles over two orders of
  magnitude. The scaling of the tree-walk is between ${\cal O}(N)$ 
  (shown with the black solid  line) and the 
  theoretical ${\cal O}(N\log N)$ and is due to the 
  average number of interactions staying roughly constant (see inset). 
  The asymptotic complexity of the tree-construction approaches ${\cal O}(N)$, 
  as expected, since all the constituent primitives share the same complexity.   
  The timings are from the GTX480 GPU with $\theta$ = 0.75.}
\label{Image:Scaling}
\end{figure}

To measure the scaling of the implemented algorithms we execute 
simulations using Plummer spheres (\cite{1915MNRAS..76..107P}) 
with $N=2^{15}$ (~32k) up to $N=2^{22}$ (~4M) particles. We measure the 
performance of the same algorithms as in the previous paragraph.
The results can be found in Fig.~\ref{Image:Scaling}.
The wall-clock time spent in the sorting, moving,
tree-construction and multipole computation algorithms scales linearly
with $N$ for $N \gtrsim 10^6$.  For smaller $N$, however, the scaling
is sub-linear, because the algorithms require more than
$10^5$ particles to saturate the GPU.  In theory the tree-traverse
scales as ${\cal O}(N\log N)$, whereas empirically the wall-clock time
scales almost linearly with $N$.  This is explained in the inset of
Fig.~\ref{Image:Scaling}, which shows the average number of
interactions per particle during the simulation.  The average number
of particle-cell interactions doubles between $N \gtrsim $~32k and 
$N \lesssim $~1M and keeps gradually increasing for $N \gtrsim $~1M.
This break is not clearly visible in the timing results since for
small particle numbers ($N \lesssim $~1M ) not all GPU resources 
are satured.
Finally, more than 90\% of the wall-clock time is
spent on tree-traversion with $\theta=0.75$. This allows for block time-step execution
where the tree-traverse time is reduced by a factor 
$N/N_{\rm active}$, where $N_{\rm active}$ is the 
number of particles that have to be updated.

\section{Discussion and Conclusions}\label{sect:discussion}

We have presented an efficient gravitational $N$-body tree-code. 
In contrast to other existing GPU tree-codes, this implementation 
is executed completely on the GPU.  On a GTX480
the number of particles processed per unit time is $2.8$ million 
particles per second with $\theta=0.75$. This allows us to routinely
carry out simulations on the GPU.
Since the current version can only use 1 GPU, the limitation
is the amount of memory. For 5 million particles $\sim 1$ gigabyte of
GPU memory is required.

Even though the sorting, moving and tree-construction parts of the
code take up roughly 10\% of the execution time in the presented
timings, these methods do not have to be executed during each
time-step when using the block time-step method.  It is sufficient to
only recompute the multipole moments of tree-cells that have updated
child particles, and only when the tree-traverse shows a considerable
decline in performance does the complete tree-structure has to be rebuild.
This decline is the result of inefficient memory reads and an increase
of the average number of particle-cell and particle-particle
interactions. This quantity increases because the tree-cell size
increases, which causes more cells to be opened by the MAC.

Although the implemented algorithms are designed for a
shared-memory architecture, they can be used to construct and traverse
tree-structures on parallel GPU clusters using the methods described
in \cite{169640, 1996NewA....1..133D}.  Furthermore, in case of a
parallel GPU tree-code, the CPU can exchange particles with the other
nodes, while the GPU is traversing the tree-structure of the local
data, making it possible to hide most of the communication
time.  

The implemented algorithms are not
limited to the evaluation of gravitational forces, but can be applied
to a variety of problems, such as neighbour search, clump finding
algorithms, fast multipole method and ray tracing. In particular, it
is straightforward to implement Smoothed Particle Hydrodynamics in
such a code, therefore having a self-gravitating particle based
hydrodynamics code implemented on the GPU.

\acknowledgements

This work is supported by NOVA and NWO grants (\#639.073.803,
\#643.000.802, and \#614.061.608, VICI \#643.200.503, VIDI
\#639.042.607). The authors would like to thank Massimiliano Fatica
and Mark Harris of NVIDIA for the help with getting the code to run on
the Fermi architecture

\bibliography{Bedorf_J}

\end{document}